\documentclass[aps, pra, reprint, amsmath, amssymb]{revtex4-1}
\usepackage{verbatim}
\usepackage{graphicx}
\usepackage{subfig}
\usepackage{epstopdf}
\usepackage{float}
\usepackage{dcolumn}
\usepackage{bm}

\begin{document}

\title{Planar waveguide structures for production of modally entangled biphotons with spectrum broadening}

\author{A.~R.~Tamazyan}
\email[]{a.tamazyan@ysu.am}
\affiliation{Yerevan State University, 1 Alex Manoogian, 0025,
Yerevan, Armenia}
\affiliation{Institute for Physical Researches, National Academy of Sciences, Ashtarak-2, 0203,\\ Ashtarak,
Armenia}
\author{\fbox{G.~Yu.~Kryuchkyan}}
\affiliation{Yerevan State University, 1 Alex Manoogian, 0025,
Yerevan, Armenia}
\affiliation{Institute for Physical Researches, National Academy of Sciences, Ashtarak-2, 0203,\\ Ashtarak,
Armenia}
\date{\today}
\begin{abstract}
We study the generation of biphoton modally entangled states in such waveguide structures that lead to spectrum broadening. The process of spontaneous parametric down conversion (SPDC) has been considered for the biphoton generation. 
The main subject of our study is the dependence of the biphoton spectra on the structure spatial characteristics. The representation of the core structures as definite assembles of nonlinear layers allows us to find analytical description for biphoton spectrum. We show that chirped biphotons with discrete as well as continuous spectra can be generated depending on the waveguide structure. The conditions for spectrum discretization are obtained analytically. Moreover it is shown that the biphoton entangled states can be controlled by varying the parameters of the waveguide structure. 
\end{abstract}


\maketitle

\section{\label{intro}Introduction}

The production of strongly correlated quantum systems is an important challenge for modern optics. Particularly generation of entangled biphotons with controllable spectral and temporal characteristics represents a major interest nowadays. It is well known that one of the most popular sources for generation of photon pairs is the process of SPDC, the process of photon splitting into two photons with lower frequencies due to interaction with a nonlinear media \cite{HsOsh,HB, MKl, MKL2, Shih}. In this interaction two conservation laws should be considered: the energy conservation law, stating that the sum of generated photon frequencies should be equal to the frequency of the pump field, and the momentum conservation law for the wave vectors of the interacting fields, or so called phase matching condition. However, due to the strong dispersion in some nonlinear media, the phase-matching condition is mostly changed to the quasi-phase matching condition(QPM), stating that  during the light propagation the phase mismatch should equal to some constant value, referred as the structure momentum. It has been shown that periodically poled nonlinear crystals as well as nonlinear photonic crystals can be used for production of biphotons, with realization of QPM conditions, depending on the structure period \cite{Kly, FrWr, Yamada, Batchko, Nakamura}.

Several issues with periodically poled structures (the low intensities of down converted light and the significant time delays), which are the consequences of narrow frequency spectrum, led to the idea for construction of new forms of nonlinear structures for the production of biphotons with wider frequency spectra and smaller correlation times at high emission rates.  One of the suggested solutions was the generation of biphotons in so called chirped QPM crystals, realized as  aperiodically poled nonlinear crystals,  having a spatial dependent poling period  unlike the periodically poled crystals. The generation of light in these structures was studied theoretically and demonstrated experimentally by several groups \cite{Crrs1,Crrs2,Crrs3, Crrs4, nast,DnUr,Harris, NsCrs,kitaeva,SenHar,MMASch}. In aperiodically poled crystals, the poling period and, as a result, the QPM parameter vary along the crystal length providing opportunity to use these structures for realization of SPDC with  a range of frequencies, thus making it possible to use the same structure as a source for down conversion of non-monochromatic wave-pocket. Down conversion in aperiodically poled crystals is mostly explained with the theory of linearly-chirped biphotons, based on the simple model of linearly changing phase-mismatch vector \cite{Imeshev,NsCrs,kitaeva,SenHar, Harris}. In this approach the phase-mismatch is given by the following formula: $k_{g}(z)=k_{0}-\alpha z$, where $k_{0}$ is the grating's spatial frequency at the  medium entrance face. With this approach the biphoton spectral amplitude can be represented with rectangular function, so the spectrum range is described by the width of this function \cite{kitaeva}. This approximation allows to explain the spectrum broadening in a simple way. However this theory is only applicable for aperiodically poled crystals with extra high number of layers with small lengths. Later it has been shown theoretically that spectrum broadening can be achieved even with smaller number of layers, using no phenomenological approximation for representation of phase-mismatch vector. The spectrum broadening can be achieved in two types of structures: aperiodically poled nonlinear crystals and nonlinear photonic crystals with linearly changing refractive index \cite {DKA}. This approach leads to the representation of the biphoton spectrum in the form of Gauss sums. It should be noted that in the periodically poled crystals the biphoton spectrum is  described as a delta-function, having a non-zero value only at the half frequency of the pump field. Controversially, both rectangular function and Gauss sums have finite width, giving a range of possible frequencies for the generated biphotons.

However the production of entangled biphotons itself is not enough for the applications in quantum engineering. The propagation of the coupled biphotons is a problem of much more value and interest. A number of applications have been engineered for the propagation of biphotons in long distances. It is not a secret that one of the most powerful instruments for light propagation without major losses are the waveguides.  
This is one of the reasons that  during the last decade major studies have been devoted to the biphoton generation in waveguides\cite{chen, beck, uren1, uren2, uren3, tanzili, christ, svozilik}. Another advantage of using waveguides as sources for biphoton generation is that a new form of entanglement  between the biphotons, so called modal entanglement, arises here in addition to polarization and frequency entanglements. It is known that in planar waveguides light propagates due to sequential reflections from the waveguide borders, and due to the border conditions, not all directions of propagation are possible, bringing to discretization of light spatial modes. When considering biphoton generation in waveguide configurations, only the longitudinal components of signal and idler photons wave vectors can be controlled in terms of realization of QPM condition, while the transverse components determine the spatial modes of photons, thus bringing to one more parameter characterizing the entanglement. It has been shown that the spatial modes of generated biphotons are connected, so the term modal entanglement can be applied to this phenomenon \cite{Saleh}. It has been shown that waveguides constructed of nonlinear media can become a source for correlated light. Waveguide structures constructed of periodically as well as aperiodically poled crystals have been considered for this purpose.

Particular interest represents the study of biphoton generation in chirped multimode waveguides. The phenomenology of the biphoton generation in chirped aperiodically poled crystals when all the three interacting fields are in the fundamental mode has been represented recently \cite{AK}.

In this paper,  we go further and focus on the study of the propagation of modally entangled biphotons in layered waveguides with two different structures. (i) Aperiodically poled layered nonlinear crystals, and (ii) layered nonlinear photonic crystals are suggested to be used as the waveguide core, leading to the widening of the biphoton spectra. SPDC type II process, where the polarizations of the generated biphotons are different,  is considered in these structures.  

 The paper  is organized as follows: in Sec. \ref{main} the chirped QPM SPDC   is described in a layered waveguide, while Sec. \ref{chirp} is devoted to the biphoton generation in waveguides with cores of different layered structures.

\section{\label{main}SPDC in waveguides: Main aspects}

It is known that SPDC can be realized due to interaction of a strong laser field with  nonlinear materials having second order nonlinear susceptibility $\chi^{(2)}$. Planar waveguides constructed of a nonlinear material such as $LaTiO_3$ or $KTP$ crystals can be used as a nonlinear medium.

In this section we discuss the light propagation in waveguides and represent the steps to obtain analytical expression for biphoton spectral amplitude. Light propagation main characteristics in multimode waveguides are represented in Sec. \ref{spatial} and the calculation of biphoton spectral amplitude is provided in Sec. \ref{spectrum}.

\subsection{\label{spatial} Spatial modes of light}

Propagation of TE waves in a generic planar waveguide  is considered here with core and cladding of refractive indexes $n_{c}$ and $n_{cl}$ respectively, with refractive index of the core bigger than of the cladding $n_{c} > n_{cl}$. In these structures light propagates in the core due to sequential reflections from the cladding surface. The border conditions between the core and the cladding bring to the discretization of light propagation directions ($\theta$ reflection angle values), and thus to the formation of the light spatial modes. Below we represent the characteristics of the light spatial modes.

The electric field amplitudes for each mode can takes the following form, \cite{Saleh}
\begin{eqnarray}
u_{\mu, \sigma} \sim \left\{ \begin{array}{ll}
cos(k_{\mu, \sigma}^{z}(\omega)z)~~|\mu=even,\\
\\
sin(k_{\mu, \sigma}^{z}(\omega)z)~~|\mu=odd.\end{array}\right.
\label{eq::amp}
\end{eqnarray}
Here, $z$ is the coordinate perpendicular to the propagation direction, $\mu$ is the spatial mode, $\omega$ is the frequency, $\sigma$ is the polarization, $u_{\mu}$ is the corresponding modal amplitude, and $k_{\mu, \sigma}^{z}$ is the z component of the wave vector.
\begin{equation}
k_{\mu}^{2} = \beta_{\mu}^{2} + {k_{\mu}^{z}}^{2}, \nonumber
\end{equation} 
where $\beta_{\mu}$ is the component of the wave vector responsible for the light propagation. For convenience in farther calculations, this quantity is represented in terms of effective refractive index,
\begin{eqnarray}
\beta_{\mu} = n_{eff}^{\mu}(\omega)\frac{\omega}{c},\\
n_{eff}^{\mu}(\omega) = n_c cos(\theta^{\mu}(\omega)), \nonumber
\label{eq::beta}
\end{eqnarray}
where $\theta$ is the  light propagation direction. Note, that the $\sigma$ notations are skipped here.

Taking into account the border conditions between the core and the cladding, the transcendent equation for the $\mu$-th mode of the light propagating through the core,  for TE waves takes the following form,
 \begin{equation}
 tan\Bigg(\frac{Hk_{\mu}^{z}}{2}-\frac{\mu\pi}{2}\Bigg)=\frac{\sqrt{(n_{c}^{2}-n_{cl}^{2})\omega^{2}/c^{2}-{k_{\mu}^{z}}^2}}{k_{\mu}^{z}}.
 \label{eq::wavevect}
 \end{equation}
Here, $H$ is the waveguide height, $n_c$  and $n_{cl}$ are the refractive indexes of the core and the cladding correspondingly, and $c$ is the speed of light.
After some trigonometric transformations, we represent  the transcendent equation in a more convenient form.
\begin{eqnarray}
cos\bigg(\frac{H n_{z, \mu}\omega }{2c}- \frac{\mu\pi}{2}\bigg)=\frac{n_{z,\mu}}{\sqrt{n_{c}^{2}-n_{cl}^{2}}}, \nonumber
\end{eqnarray}
and finally, for 0 and 1 modes correspondingly,
\begin{eqnarray}
cosc\bigg(\frac{H n_{z, 0}\omega }{2c}\bigg)=\frac{2c}{H\omega\sqrt{n_{c}^{2}-n_{cl}^{2}}},
~~~\nonumber \\
sinc\bigg(\frac{H n_{z, 1}\omega }{2c}\bigg)=\frac{2c}{H\omega\sqrt{n_{c}^{2}-n_{cl}^{2}}},
~~~
\label{eq::trfinal}
\end{eqnarray}
with notations $cosc(x)= \frac{cos(x)}{x}$, $sinc(x)=\frac{sin(x)}{x}$, and $n_{z, \mu} = \sqrt{n_{c}^{2}-{n_{eff}^{\mu}}^{2}}$.

\begin{figure}
\includegraphics[width=7cm]{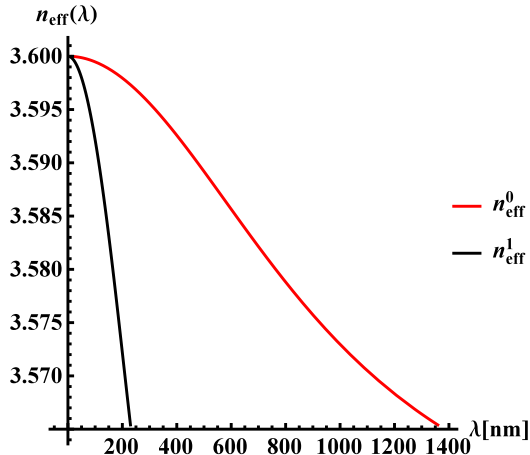}
\caption{Effective refractive indices for 0 and 1 modes in dependence on the wavelength, $H= 0.8 \mu m$, $n_{cl} = 3.5$ and $n_c = 3.6$.} 
 \label{fig::n_eff}
\end{figure}

Equation \ref{eq::trfinal} can be used for the calculation of the effective refractive index numerically in waveguides with different core and cladding configurations. These generic results allow to analyze a range of problems concerning the light propagation in waveguides. Particularly we use these equations to obtain the characteristics of biphoton generation in waveguides with different structures. 

Figure \ref{fig::n_eff} shows numeric results for 0 and 1 mode effective refractive index dependence on the wavelength, for $LaTiO_3$ crystal used as the waveguide core (refractive index ~3.6), the cladding refractive index is taken to be 3.5. The results represented in this figure show that there are wavelength limitations for light propagation in each mode. These parameters are used in all our farther calculations.

\begin{figure*}
\begin{center}
\subfloat[]{\includegraphics[width=8cm]{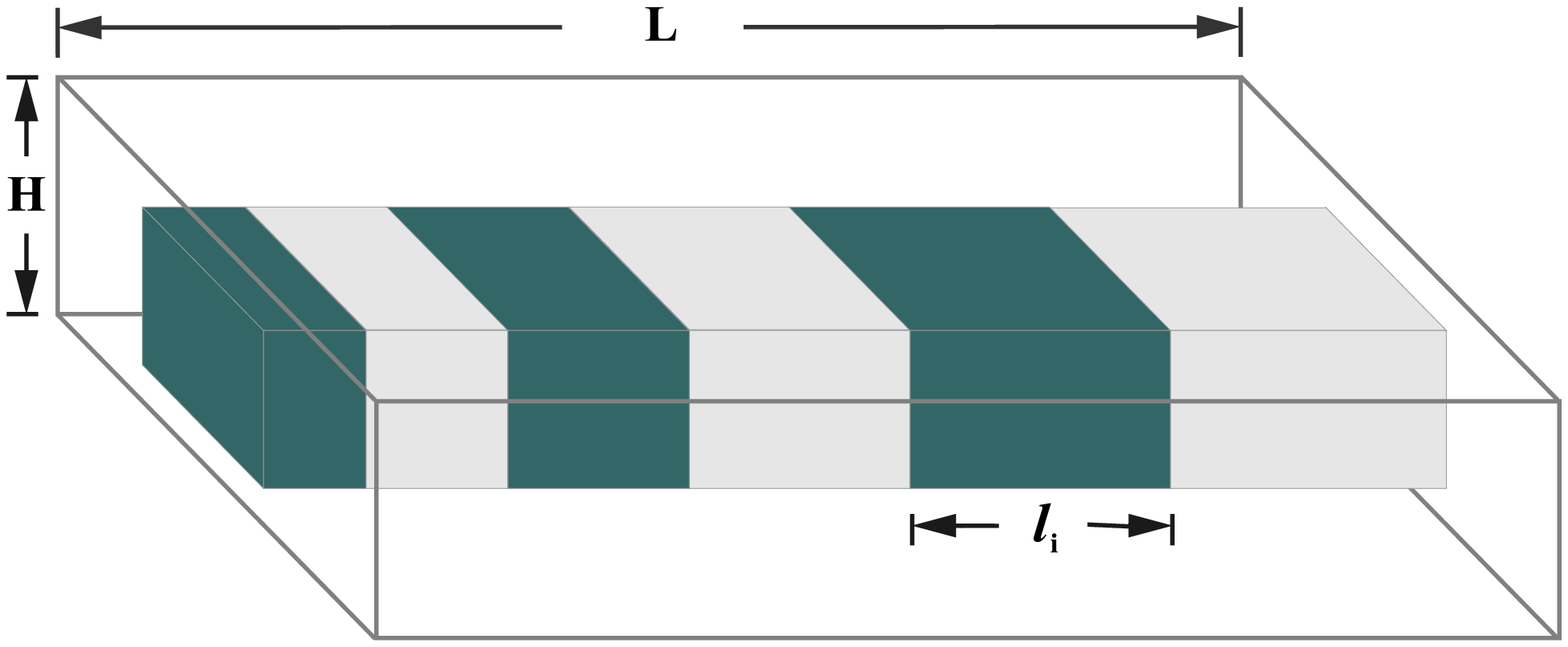} \label{fig::layered_a}}%
\qquad
\subfloat[]{\includegraphics[width=8cm]{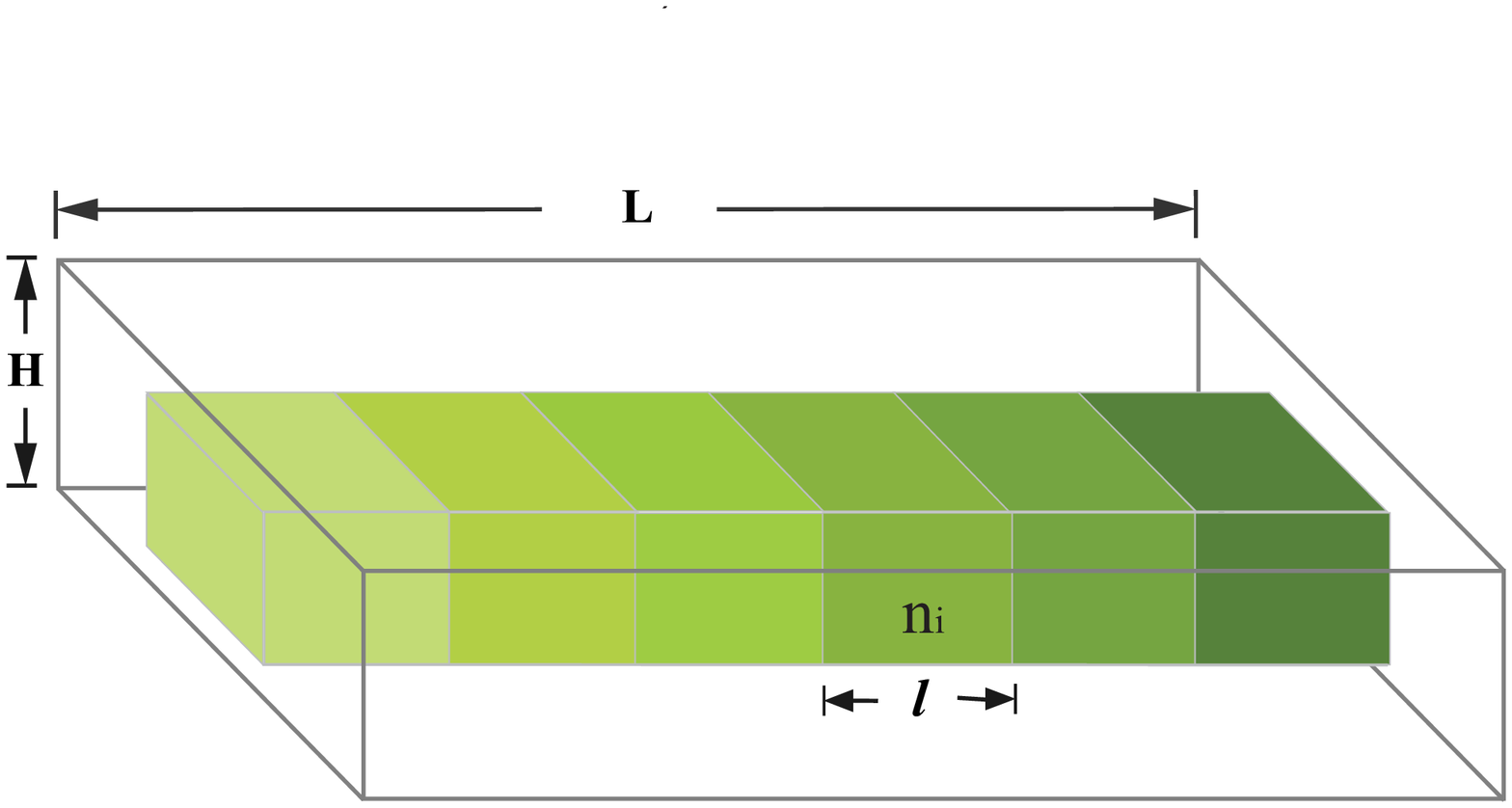} \label{fig::layered_b}}%
\caption{ Planar waveguide structures.
(a) aperiodically poled layered waveguide, dark colored layers correspond to positive nonlinear susceptibility and light colored layers to negative.
(b) photonic crystal like waveguide, the change in layer coloring represents the refractive index change in the layers.}
 \label{fig::layered}
 \end{center}
\end{figure*}

\subsection{\label{spectrum} Biphoton spectrum}

Three photon interaction in a nonlinear media with second order susceptibility $\chi^{(2)}$ can be described via the following semi-classical Hamiltonian: 

 \begin{equation}
\hat{ H}(t)=\int_{0}^{V}{d\vec{r}\chi^{(2)}(\vec{r})E_{p}^{*}(\vec{r},t)\hat{E_{i}}^{(-)}(\vec{r},t)\hat{E_{s}}^{(-)}(\vec{r},t)+h.c.}
\label{eq::hamiltonian}
 \end{equation}

Here, the integration is performed over the medium volume V, $\chi^{2}(\vec{r})$ is the second order nonlinear susceptibility of the medium that in general is spatial dependent, $E_p(\vec{r},t)$ is the amplitude of the pump field, while $E_{i}^{+/-}(\vec{r},t)$ are the operators of the idler, and  $E_{s}^{+/-}(\vec{r},t)$ of the signal fields.

For the polarized and multimode light in a waveguide the electric-field operators of the signal and idler waves propagating along the x direction can be represented as

\begin{eqnarray}
\hat{E}_{q}^{(-)}(\vec{r}, t)=\int d\omega_{q}\sum_{\mu,\sigma}\hat{a}_{\mu,\sigma}^{+}(\omega_{q})u_{\mu,\sigma}
 e^{i(\omega_{q}t-\beta_{\mu,\sigma}x)},
\label{eq::subh}
\end{eqnarray}
where $\omega_{q}$ is the frequency, $\mu$ and $\sigma$  represent the mode and polarization of the light respectively, $\hat{a}^{+}$ is the creation operator of the corresponding field and $\beta$ is the projection of the corresponding wave vector on x axis, defined by Eq. \ref{eq::beta}.

The amplitude of a monochromatic pump field can be represented in a simple classical formula,

\begin{eqnarray}
E_{p}^{*}(\vec{r}, t)\sim u_{p}e^{-i(\omega_{p}t-\beta_{p}x)},
\label{eq::pump}
\end{eqnarray}
$\omega_{p}$ is the frequency of the pump field. 

From the energy conservation law the following relation between the pump and subharmonic field frequencies arises $\omega_{p}=\omega_{s}+\omega_{i}$.

Next we turn to the representation of biphoton generation in layered structures.
After assumption that media characteristics do not change within a single layer, interaction Hamiltonian  in a layered structure can be represented as a superposition of interactions within the layers
\begin{equation}
 \hat{H}(t)=\sum_{m}\hat{H}_{m}(t), \label{eq::h_l}
 \end{equation}
 where  $\hat{H}_{m}(t)$ is the Hamiltonian within m-th layer given by Eq. \ref{eq::hamiltonian}, integrated over the layer's space.

Using the perturbation theory, the two photon state can be represented via the interaction Hamiltonian in the following form,
\begin{equation}
|\Psi\rangle \sim \int dt\hat{H}(t)|0,0\rangle.
\end{equation} 

After placing the corresponding expression for Hamiltonian given by Eq. \ref{eq::h_l} and applying Fourier transformation, this representation of the state function can be replaced with the integral via frequencies.
\begin{equation}
|\Psi\rangle \sim \int d\omega_{s}\sum_{\mu,\sigma} \Phi_{\mu,\sigma}(\omega_s)|\omega_{s},\sigma_{s}\rangle |\omega_{i}\sigma_{i}\rangle,
\end{equation}
where, $\Phi_{\mu, \sigma}$ denotes the spectral amplitude of the biphotons. The analysis of this function gives the full image of biphoton spectra.

 Next, the spectral amplitude actually represents a sum over the layers, and thus can be represented as the superposition of spectral amplitudes within each layer:
\begin{eqnarray}
\Phi_{\mu, \sigma}(\omega_{s}) = \sum_{m}\Phi_{m,\mu, \sigma}(\omega_{s}), \label{eq::Phi}\\
\Phi_{m,\mu,\sigma}(\omega_{s})=A_{m, \mu,\sigma}(\omega_{s})\int_{x_{m}}^{x_{m+1}} dx \chi^{(2)}_{m}e^{i\Delta\beta_{m}x},\\
A_{m, \mu,\sigma}(\omega_s)=\int\int dy dz \prod_{q =p, s,i} u_{m,\mu, \sigma}^{q}(\omega,y,z),\label{eq::APhi}
\end{eqnarray}
and  after integration, the following formula is obtained for the spectral amplitude in a single layer,
\begin{eqnarray}
\Phi_{m, \mu, \sigma}= l_{m}\chi^{(2)}_{m}A_{m,\mu,\sigma}(\omega_{s})\times \nonumber ~~~~~~~~~~~~~~~~~~~~~~~~~~\\
\times\exp(-i(\phi_{m}+\frac{\Delta \beta_{m}l_{m}}{2}))
 sinc\Bigg(\frac{\Delta\beta_{m}l_{m}}{2}\Bigg).
\label{eq::Phi_m}
\end{eqnarray}
Here $A_{m, \mu, \sigma}$ is the spatial part of the spectral amplitude, further referred as spatial amplitude, and $\phi_m$ is the following phase
\begin{equation}
\phi_{m}=\sum_{n}^{m - 1}\Delta \beta_{n}l_n. \nonumber
\end{equation}

It follows from Eq. \ref{eq::APhi}, that SPDC process in multimode waveguides is possible not between all modal combinations of the pump, signal and idler fields. If the production of three amplitudes is an odd function, the spatial amplitude, and therefore, the spectral amplitude vanishes. This means that the number of odd modes in the wave triplet (p, s, i) should be even. For an example, if the pump field propagates in mode 1 and the signal photon is in mode 0, the idler photon can propagate in mode 1, or any other odd mode.

Next, spatial amplitudes for the three waves propagating in  particular modes should be calculated.The spatial amplitude  $A_m$ can be found using Eqs. \ref{eq::amp} and \ref{eq::APhi}.

Let us consider the case when all three waves propagate in fundamental 0 mode.
We do not consider any inhomogeneity  in y direction, so the integral over y axis will only bring to a constant, and the integral over the z axis can be calculated easily,

\begin{eqnarray}
A_{m, \mu}=A_{0}\int_{-H/2}^{H/2}\frac{ 2dz}{H}sin(k^{z}_{p,m,\mu}z) sin(k^{z}_{s,m,\mu}z) cos(k^{z}_{i,m,\mu}z), \nonumber\\
A_{m, \mu}=A_{0}\Bigg[sinc\Bigg(\frac{(k_{p,m,\mu}^{z}+k_{s,m,\mu}^{z}+k_{i,m,\mu}^{z})H}{2}\Bigg)-~~~~~~~~~~~\nonumber \\
-sinc\Bigg(\frac{(k_{p,m,\mu}^{z}-k_{s,m,\mu}^{z}-k_{i,m,\mu}^{z})H}{2}\Bigg) +~~~~~~~~\nonumber \\
+sinc\Bigg(\frac{(k_{p,m,\mu}^{z}+k_{s,m,\mu}^{z}-k_{i,m,\mu}^{z})H}{2}\Bigg)-~~~~~~~~\nonumber \\
-sinc\Bigg(\frac{(k_{p,m,\mu}^{z}-k_{s,m,\mu}^{z}+k_{i,m,\mu}^{z})H}{2}\Bigg)\Bigg].~~~~~~~~
\label{eq::A_m}
\end{eqnarray}

Here again  the $\sigma$ notations are skipped, $A_{0}=\frac{u_{0p}u_{0s}u_{0i}L_{y}H}{2}$ is a constant that describes the height of the spectrum, $L_y$ is the width of the core, $H$ is the height of the core, m is the layer number, $k_ {q, m}^{z}$ ($q = p, s, i$) denote the z components of the spatial amplitudes of the three waves, pump, signal and idler correspondingly. 

The spatial amplitude for the modal configuration  when the pump and idler(signal) fields are in mode 1 and the signal(idler) is in mode 0, can be found in a similar way.

The value of  $A_{m,\mu}$ spatial amplitude does not change with the layers, as from the boundary conditions between the layers it follows that the $k_{q, m}^z$ component of the spatial frequency should maintain its value during the propagation. Equation \ref{eq::A_m} in combination with Eqs. \ref{eq::Phi} and \ref{eq::Phi_m} represents a generic formula for calculation of the biphoton spectra in  layered waveguides.

\begin{figure}
\centering
\subfloat[]{\includegraphics[width=8cm]{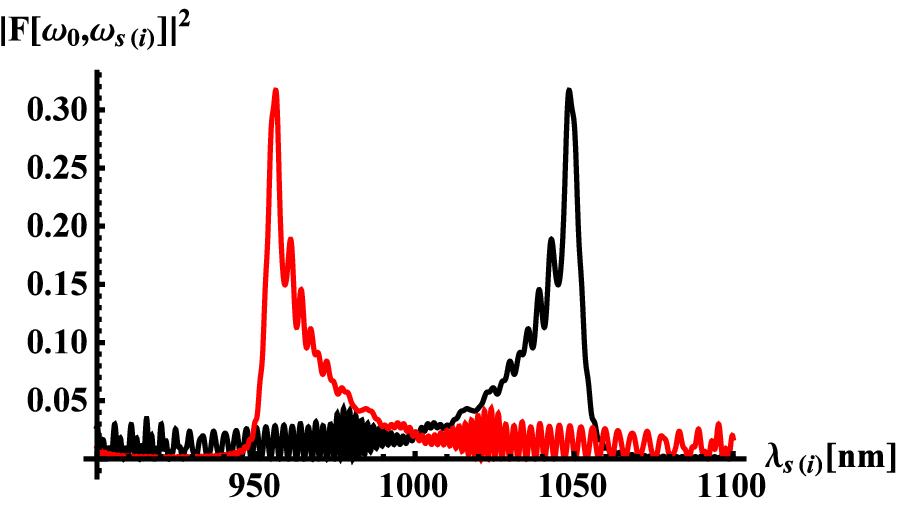} \label{fig::Spe_a}}%
\qquad
\subfloat[]{\includegraphics[width=8cm]{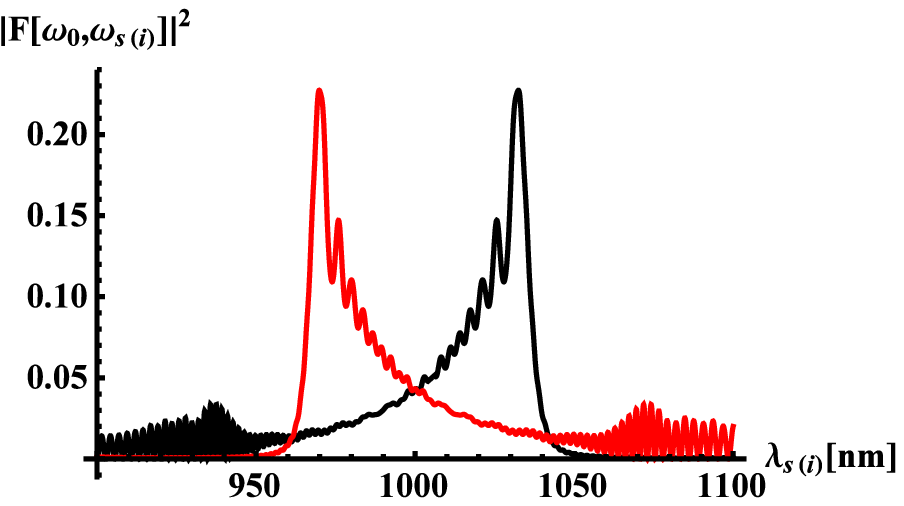} \label{fig::Spe_b}}
\caption{Modal entanglement of biphotons in aperiodically poled waveguides for waveguide parameters:  $L=12000 \mu m$, chirp
parameter $\beta = 0.25\mu m (a)$, $\beta = 0.5\mu m (b)$, $H= 1 \mu m$, and with number of layers $N = 100$. Spectral amplitudes are depicted for signal and idler photon states $|\mu_{s}, \mu_{i} \rangle = |0,1\rangle $ (black bold curve) and $|\mu_{s}, \mu_{i}\rangle = |1, 0\rangle$ (red bold curve).} 
 \label{fig::Spe}
\end{figure}

\section{\label{chirp}Applying chirp to multimode biphoton production}

In this section we turn to the calculation and analysis of biphoton spectra for particular structures. Two different models of layered structure are suggested  to obtain biphoton spectrum widening: aperiodically poled crystals and periodically poled photonic crystals with refractive index changing  in the layers (Fig. \ref{fig::layered}).

\subsection{\label{aperiodic}Aperiodically poled crystals}

A simple model for aperiodically poled crystals is a layered nonlinear structure with increasing layer length in the light propagation direction (x), and the second order nonlinear susceptibility  changing its sign in each next layer (Fig. \ref{fig::layered_a}),
\begin{eqnarray}
l_{m}= l_{0} + m\varsigma,\label{eq::l_m}\\
\chi^{(2)}_m = (-1)^{m}\chi_{0}^{(2)},\nonumber
\end{eqnarray}
where $l_m$ is the length of m-th layer, $\varsigma$ is the difference of neighboring layer lengths, which plays the role of chirp parameter for these structures, and $\chi^{(2)}_m$ is the second order nonlinear susceptibility in the m-th layer.

The biphoton spectral amplitude for this structure can be obtained using the generic formula given by Eq. \ref{eq::Phi_m}:

\begin{widetext}
\begin{eqnarray}
|\Phi(\lambda_{s})|^{2}= \frac{A_{\mu}^2{\chi_{0}^{(2)}}^2}{\Delta\beta^2}\sum_{m=1}^{N}\Bigg[sin^{2}\Bigg(\frac{\Delta\beta l_{m}}{2}\Bigg) + 2\sum_{p=1}^{N-m}(-1)^{p}sin\Bigg(\frac{\Delta\beta l_{m}}{2}\Bigg)sin\Bigg(\frac{\Delta\beta l_{m+p}}{2}\Bigg)cos(\zeta_{m,p})\Bigg].
\label{eq::SpAprd}
\end{eqnarray}
\end{widetext}
Here  $\zeta_{m,p}= \Delta\beta p(l_{0}+(m+p/2-1)\varsigma)$, $\lambda_s$ is the wavelength of the signal field, $\Delta\beta$ is the phase mismatch in the propagation direction, $l_m$ is the length of m-th layer, and $\varsigma$ is the chirp parameter given in Eq. \ref{eq::l_m}.

Next, biphotons are considered to have frequency values in the same region, so one can expand the wave vectors of the subharmonics in the Taylor series near $\omega_p/2$. Given the second-order expansion, we have as a result:
\begin{equation}
\Delta\beta_m(\Omega) = \Delta\beta_{m0}+D\Omega+B\Omega,
\end{equation}

where $\Delta\beta_{m0}=\beta_{s,m}(\omega_p/2)+\beta_{i, m}(\omega_p/2)-\beta_{p,m}(\omega_p)$ is the phase mismatch at the biphoton central frequency, which is usually shortened with the QPM parameter for the case of reoriented crystals. 
$D=\beta^{\prime}_{s,m}(\omega_p/2)-\beta^{\prime}_{i,m}(\omega_p/2)$ is the time delay between the subharmonic modes, and $\Omega = \omega_s-\omega_p/2=\omega_p/2-\omega_i$ determines the frequency difference between idler and signal fields.

Now we can turn to the  analysis of the modal entanglement of  biphotons. The pump field is considered to propagate in mode 1, hence  leading to the idler photon propagation in mode 0(1) and the signal photon in mode 1(0). 

The biphoton state function for modally entangled states can be written as
\begin{eqnarray}
|\Psi\rangle \sim \int d\omega_s(\Phi_{0,1}|\omega_s,0\rangle|\omega_i,1\rangle + \Phi_{1,0}|\omega_s,1\rangle|\omega_i,0\rangle).
\end{eqnarray}
Thus, the condition for the biphoton modal entanglement can be stated as the condition of $\Phi_{0,1}$ and $\Phi_{1,0}$ coexistence.
The biphoton spectrum is obtained using Eq. \ref{eq::SpAprd}, and the effective refractive index value for the corresponding center frequency (see Fig. \ref{fig::n_eff}). The results are illustrated in Fig. \ref{fig::Spe}. The figure shows biphoton spectra widening for these structures. However, it can be noticed that the spectra is not symmetric, in terms that the amplitude values, and thus the intensities are different for different frequency values of the subharmonics. It can also be noticed that there is an overlap between the (0, 1) and (1, 0) spectra. This overlapped area corresponds to the frequency region where the biphotons are entangled. Comparison between Figs. \ref{fig::Spe_a} and \ref{fig::Spe_b} shows that the spectrum width as well as the entanglement region can be controlled by varying the waveguide structure, particularly the chirp parameter and the layer lengths.

\subsection{\label{photonic}Photonic crystals}

The other structure considered for biphoton generation is a layered structure with layers of same lenght, $l_{m}=l$, and with second order susceptibility changing its sign with the layers $\chi^{(2)}_{m}=(-1)^{m}\chi^{(2)}_{0}$, while refractive indexes of the pump
 and the generated waves $n_{c}(\omega_{q},z)=n_{c}^{q}(x)$ ($q=p,s,i$) change linearly with the layers (Fig. \ref{fig::layered_b}) and are given by
\begin{eqnarray}
n_{c}^{q}(x)=\left\{ \begin{array}{ll}
n_{c}^{q}, & ~ 0<z<l,\\
\\
n_{c}^{q}+ml\varsigma_{q}, & ~ ml<z<(m+1)l,\end{array}\right.\nonumber
\end{eqnarray}
where  $\varsigma_{q}$ ($q=p,s,i$)is the chirp parameter for the refractive indexes of the pump and subharmonic fields correspondingly. In this case, the phase-mismatch vector for the $m$-th layer is written as $\Delta \beta_{m}=\Delta \beta-\alpha( m-1)l$, where $\Delta\beta=\beta_{0}(\omega_{0})-\beta_{s}(\omega_{s})-\beta_{i}(\omega_{i})$ is the x axis projection of the phase mismatch vector at the first layer, and $\alpha=\frac{\omega_{0}}{c}(\varsigma_{0}-\frac{\varsigma_{s}}{2}-\frac{\varsigma_{i}}{2})$ is the spatial chirp parameter. This structure makes a convenient media for generation of chirped biphotons \cite{DKA}.

\begin{figure}
\begin{center}
\subfloat[]{\includegraphics[width = 8cm]{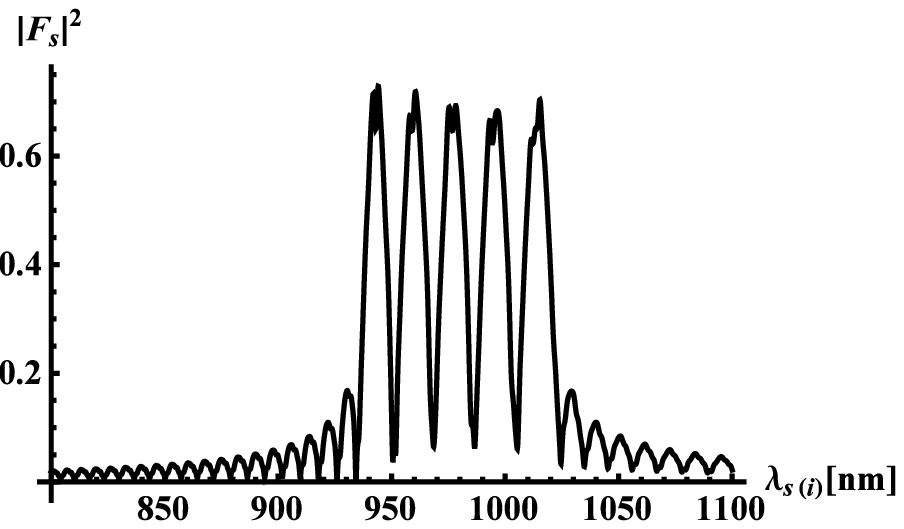}\label{fig::rindex_a}}
\qquad
\subfloat[]{\includegraphics[width= 8cm]{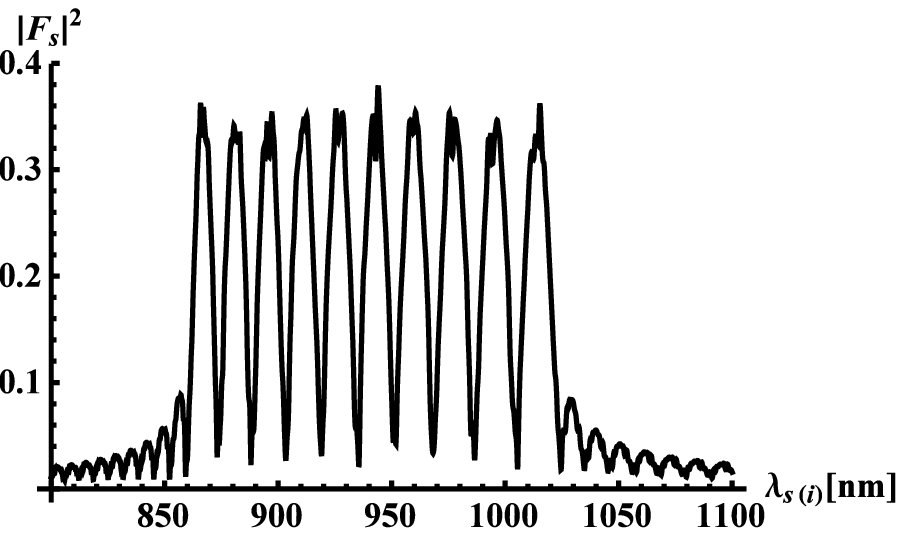}\label{fig::rindex_b}}
\qquad
\subfloat[]{\includegraphics[width = 8cm]{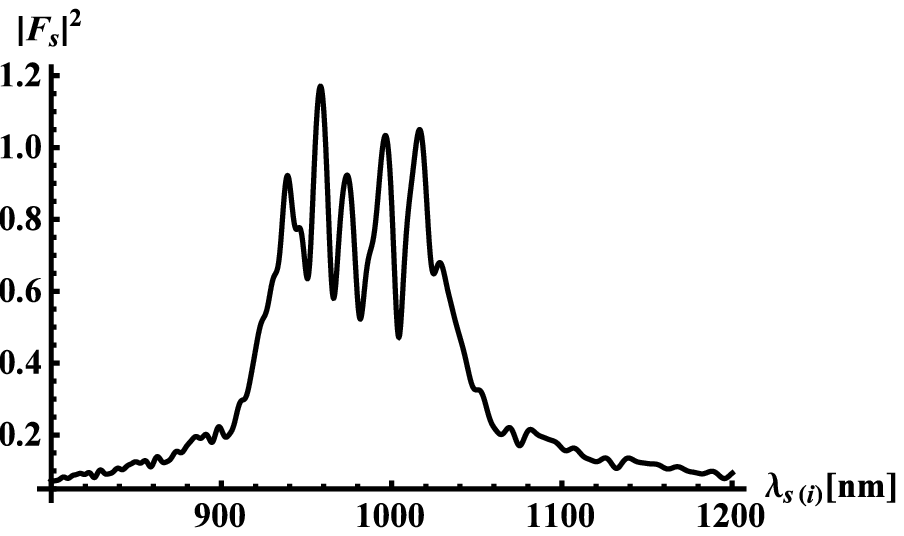}\label{fig::rindex_c}}
\caption{Biphoton spectral amplitude in photonic crystal like waveguides with different characteristics. $\alpha = 2.6\times10^{-5}$, $L = 2300 \mu m$, $N=5$, $n = 2$ (a), $\alpha = 2.6\times10^{-5}$, $L = 4600 \mu m$, $N=10$, $n = 2$ (b), $\alpha = 2.6\times10^{-5}$, $L = 480 \mu m$, $N=5$, $n = 3$ (c).}
\label{fig::rindex}

\end{center}
\end{figure}
 
However, for the consideration of light propagation in waveguides, only having a structured core with the described properties is not enough, as the boundary conditions should as well be satisfied. This means that if the refractive index in the core changes linearly, the refractive index of the cladding should change in a way that allows the spatial modes not to change their properties within the layers.

It follows from the boundary conditions between the layers, that for the light to propagate, the $k_{z}$ part of the wave vector should not change with the layers. And from the transcendent equations it can be found that for the light not to change its modal characteristics with the layers, the cladding should also be layered, with refractive index changing with the layers in a  more complicated way:
\begin{eqnarray}
{n_{cl}^{q}}_{m}^{2} -{n_{cl}^{q}}_{0}^{2} = {n_{c}^{q}}_{m}^{2} -{n_{c}^{q}}_{0}^{2}~~~~~~~~~~~~~~\nonumber
\\{n_{cl}^{q}}_{m} = \sqrt{{n_{cl}^{q}}_{0}^{2} + ml\varsigma_{q}(2{n_{c}^{q}}_{0} + ml\varsigma_{q})},
\end{eqnarray}
where, ${n_{cl}}_{m}$ denotes the refractive index in the m-th layer of the cladding, and ${n_c}_{m}$ is the refractive index in the m-th layer of the core.

The biphoton spectral amplitude in photonic crystals can be then calculated, using the Eq. \ref{eq::Phi_m},
\begin{widetext}
\begin{eqnarray}
|\Phi(\lambda_s)|^{2}= A_{\mu}^2 l^2{\chi_{0}^{(2)}}^2\sum_{m=1}^{N}\Bigg[sinc^{2}\Bigg(\frac{\Delta\beta_{m}l}{2}\Bigg)
+2\sum_{p=1}^{N-m} sinc\Bigg(\frac{\Delta\beta_{m}l}{2}\Bigg) sinc\Bigg(\frac{\Delta\beta_{m+p}l)}{2}\Bigg)cos(\zeta_{m,p})\Bigg].
\end{eqnarray}
\end{widetext}

Here $\Delta\beta_{m} = B\Omega+\alpha(m-1)l$, $\zeta_{m,p} = pl(B\Omega+\alpha p(m-\frac{p}{2}-1)l)$, $\Omega=\omega_{s}-\frac{\omega_{p}}{2} =\frac{\omega_{p}}{2}-\omega_{i}$, N is the number of layers, and B is the dispersion parameter.

 The QPM condition can be realized for different layers, allowing the QMP parameter to take variety of values. It can be shown easily that waveguide parameters determining the index of the layer for which the quasi phase matching condition is satisfied, are the length of a single layer and the chirp parameter:
 \begin{equation}
 \Delta\beta(\frac{\omega_p}{2}) = \pi/l + (n-1)\alpha l,
 \end{equation}
where $n$ can take values from 1 to $N$.
This property of the structure makes the generation of biphotons with non-monochromatic pump fields, as well as the usage of the same structure for the generation of biphotons with different frequencies possible.

Another important property of these structures is that the biphoton spectra here can take discrete form depending on the chirp parameter and the layer length (see Fig. \ref{fig::rindex}). As the maxima of $sinc$ function is at point 0, hence fore the peak within each layer is achieved at $\Delta\beta_{m} = 0$. The condition for discrete spectrum is that the peaks are allocated far enough from one another, so the distances between them are much greater than the width of $sinc$ function. This also means that for the structures where  discrete spectrum is achieved, the number of the peaks always equals to the number of the layers. 

 It is known that $sinc$ function can be approximated to the gauss function $e^{-\gamma x^2}, \gamma=0.189$, and the value of this function reduces $e$ times at  $\pm\frac{1}{\sqrt{\gamma}}$. So the discrete spectrum condition can be represented as

\begin{eqnarray}
\Omega_{m} -\Omega_{m-1} = \frac{\alpha l}{2B}>>\frac{2}{Bl\sqrt{\gamma}}, \nonumber \\
\alpha l^{2} >>4/\sqrt{\gamma}.\label{eq::discrete}
\end{eqnarray}

\begin{figure}
\subfloat[]{\includegraphics[width = 8cm]{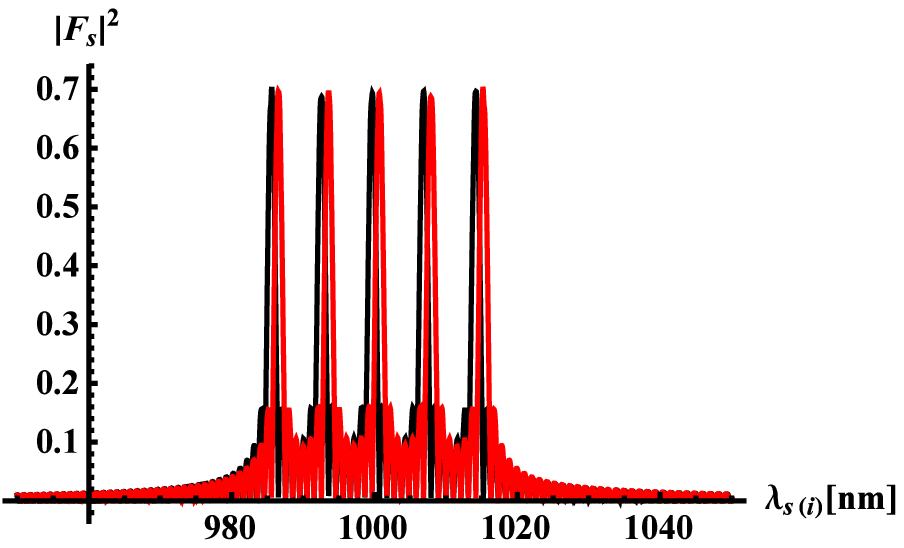} \label{fig::ent_rindex_a}}
\qquad
\subfloat[]{\includegraphics[width = 8cm]{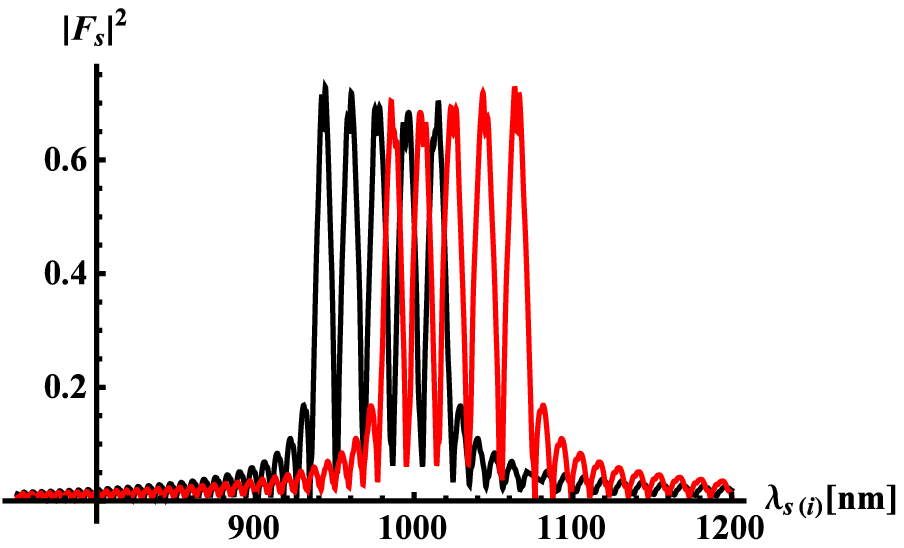}\label{fig::ent_rindex_b}}
\caption{Modal entanglement in photonic crystal like waveguides with the following parameters.  $\alpha = 1\times10^{-5}$, $L = 2000 \mu m$, $N=5$, $n = 3$ (a), $\alpha = 2.6\times10^{-5}$, $L = 2000 \mu m$, $N=5$, $n = 2$ (b), spectral amplitudes are depicted for signal and idler photon states $|\mu_{s}, \mu_{i} \rangle = |0,1\rangle $ (black bold curve) and $|\mu_{s}, \mu_{i}\rangle = |1, 0\rangle$ (red bold curve).}
 \label{fig::ent_rindex}
\end{figure}

We now turn to the study of characteristics of biphoton modal entanglement in these structures. Similar to the case of aperiodically poled crystals we analyze the 3 wave configurations where the generated biphotons are in different spacial modes (signal photon in mode 1(0) and idler in mode 0(1)). The results are depicted in Fig. \ref{fig::ent_rindex}. The graphics in Figs. \ref{fig::ent_rindex_a} and \ref{fig::ent_rindex_b} show that the entanglement region can be controlled by changing the structure characteristics. it is possible to configure structures without any entanglement of biphotons, or structures for the biphotons with fully entangled spectra simply by choosing right configurations between the layer length and the chirp parameter. This feature can be used in the sphere of quantum information where the propagation of entangled biphotons in long distances is one of the most important problems to solve.

\section{Conclusion}

The usage of waveguides as sources for generation of entangled biphotons  makes possible to propagate entangled light on long distances. This unique opportunity makes the description of biphoton generation and propagation in the waveguides valuable for a variety of applications, specifically  in the fields of quantum teleportation and information.

We have shown that waveguides with simple layered structures can be used for the generation of entangled biphotons with wide spectrum.
The main focus of this research was the study of the entangled biphoton states in waveguides of specific core and cladding structures. The spatial and spectral characteristics of biphoton propagation in waveguides have been represented, using the basics of waveguide theory and the nonlinear optics.  A generic expression describing the biphoton spectrum in layered waveguides has been obtained, allowing detailed description of the biphoton spectra in waveguides with different structures. The formation of entanglement between the spatial modes of the generated photons has been considered.

Two structures for a waveguide core and the cladding have been suggested for the generation of entangled biphotons with wide spectrum: aperiodically poled layered nonlinear crystals and periodically poled layered nonlinear photonic crystals with refractive indexes changing with the layers.
The analysis of biphoton spectrum dependence on the waveguide structure characteristics has been represented. It has been theoretically shown that biphotons with discrete as well as continuous spectrum are realizable depending on the waveguide parameters, more specifically the chirp parameter and the layer length, the theoretical limit between these parameters has been obtained for discretization of the biphoton spectrum.

At last, the spectra of modally entangled biphotons  have been analyzed. The modal configurations of generated biphotons when the pump field propagates in mode 1 have been considered.  Moreover it has been shown that for various waveguide structures the entanglement frequency regions can be changed, allowing to generate entangled biphotons across the whole spectrum or just in a narrow range of frequencies. 

\section*{Acknowledgments}
We acknowledge the support of the Armenian State Committee of Science, the Project No.15T-1C052 and EC for the RISE, Project CoExAN GA644076.

\providecommand{\noopsort}[1]{}\providecommand{\singleletter}[1]{#1}%

\end{document}